\documentstyle[12pt]{article}
\newcommand{\bea}{\begin{eqnarray}}
\newcommand{\eea}{\end{eqnarray}}
\newcommand{\bear}{\begin{eqnarray*}}
\newcommand{\eear}{\end{eqnarray*}}

\begin{document}
\begin{center}
{\bf EXACT SOLUTION OF THE BIQUADRATIC SPIN-1  t-J MODEL IN ONE 
DIMENSION }
\end{center}
\begin{center}
{F.C.Alcaraz and R.Z.Bariev\footnote{Permanent address: The Kazan Physico-Technical Institute of the Russian Academy of Sciences, 
Kazan 420029, Russia}}
\end {center}
\begin{center}
{Departamento de F\'{\i}sica, 
Universidade Federal de S\~ao Carlos, 13565-905, S\~ao Carlos, SP
Brasil}

\end{center}
\vspace{1cm}
\begin{center}

\vspace{1cm}

\today
\end{center}
\begin{abstract}
A new generalization of the t-J model with a 
nearest-neigbor hopping is formulated and solved 
exactly by the Bethe-ansatz method. The model describes the dynamics of 
spin-S fermions with isotropic or anisotropic interactions. In the case  
S=1  the magnetic interaction is biquadratic in the spin operators. 
In contrast to the $SU(N)$ 
generalization of the t-J model, studied previously in the literature, 
the present model possesses beyond a massless excitation also a massive one. 
The physical properties indicate the existence of 
Cooper-type pairs with finite binding energy.
\end{abstract}
\newpage
The t-J model has emerged as a paradigm for studying the low-energy 
electronic properties of the copper-oxide-based high-temperature 
superconductors [1,2]. 
 Although high-$T_c$ 
cuprates are at least two-dimensional systems the one-dimensional version 
of the model and its generalizations are also intensively studied 
since in this case exact results can be derived [3-12].  
  The t-J model 
describes the dynamics of spin-$\frac{1}{2}$ fermionic particles with 
Hamiltonian given by
\bea
H = -t \sum_{j,\sigma}P\left(c_{j,\sigma}^+c_{j+1,\sigma} + 
c_{j+1,\sigma}^+ c_{j,\sigma}
 \right)P + J\sum_j \left(\vec{S_j}. \vec{S}_{j+1}
 - n_j n_{j+1}/4  \right),
\eea
where $c_{j,\sigma}$ is the standard fermion creation operator, 
$\vec{S_j} = \frac{1}{2}\vec{\sigma_j}$ is the particle-spin operator and 
$n_j$ is the particle-number operator at site j. The projection operator 
$P$ excludes the double occupation at each site. 
Unfortunately the exact integrability of (1) is obtained only at 
the supersymmetric point $J = 2t$ [3-6]. At this point the model has 
no gap and the critical exponents governing the long-distance behavior of 
 correlation functions were calculated [7]. These results show that 
for any density of holes the spin-spin correlation functions dominate the 
superconducting ones, and as a consequence the model has no superconducting 
properties. These results were extended to the $SU(N)$ generalization of the
$t-J$ model  of fermions of arbitrary spin $S$ [8 - 10]. The integrability of 
an anisotropic generalization of the $SU(N)$ supersymmetric $t-J$ model has 
been shown [13, 14] and the critical exponents of the correlation functions
have been calculated [15, 16].

In this letter we present a new set of models of strong-correlated electrons 
which are exactly solvable. The first example of these models is the spin-1 
biquadratic t-J model with Hamiltonian given by
\bea
H = -t \sum_{j,\sigma}P\left(c_{j,\sigma}^+c_{j+1,\sigma} + 
c_{j+1,\sigma}^+ c_{j,\sigma} \right)P - J\sum_j \left[\left(\vec{S_j}. 
\vec{S}_{j+1}\right)^2 
 - n_j n_{j+1} \right],
\eea
where now $\vec{S_j}= (S_j^x, S_j^y, S_j^z)$ are spin-1 Pauli operators 
located at site $j$. We show that this model is exactly integrable at the 
special point $t = J$. Actually the above Hamiltonian is the isotropic 
version of a new family of anisotropic models describing the 
dynamics of spin-$S$ fermions with Hamiltonian 
\newpage
\bea
H &=& - \sum_{j=1}^L \sum_{s=-S}^{S} P\left(c_{j,s}^+c_{j+1,s} + 
c_{j+1,s}^+ c_{j,s} \right)P \nonumber\\
&-& \varepsilon \sum_{j=1}^L \left[\sum_{s,t=-S}^{S} 
u_s u_t c_{j,s}^+ c_{j,t} 
c_{j+1,-s}^+ c_{j+1,-t} - (1+\varepsilon_1){\cosh\gamma} n_j n_{j+1}\right]
\eea
where $L$ is the lattice size, 
$\varepsilon, \varepsilon_1 = \pm 1$ and the 
parameters $u_s$, which play the role of anisotropies should satisfy 
$u_s = 1/u_{-s}$ $(s = -S, -S +1, \ldots, S)$ and $ 2\cosh\gamma = u_{-S}^2 + 
u_{-S + 1}^2 + ... + u_S^2 $. The particular case  
$S = \frac{1}{2}$ 
and ${\varepsilon}=-\varepsilon_1 = 1$ 
is the anisotropic version of the supersymmetric t-J 
model. The biquadratic t-J model, at t=J, given in (2) is obtained by 
choosing in (3) $S=1$, 
${\varepsilon}=-\varepsilon_1 = 1$ and 
$u_{-1} = u_0 = u_1 = 1$. For general spin $S$ the magnetic interactions 
can be written as a polynomial of degree $2S$ in the spin operators.

The exact integrability of these models, from 
a mathematical point of view, comes from the fact that the Hamiltonian 
density in (3) is related to the generators of Hecke algebras [17], 
with deformation parameter 
 $q$ given by the relation 
$q + 1/q = 2\cosh\gamma$.

The  eigenstates and eigenvalues of 
Hamiltonian (3) can be obtained exactly within the framework of the 
Bethe-ansatz method [18-21]. The structure of the 
Bethe-ansatz equations follows from the solution of the two-particle 
problem. The two-electron wave function can be written as a 
product of two factors: a coordinate wave function (referring to the 
positions and momenta of the particles) and a spin part, the global 
wave function being antisymmetric under the exchange of two 
particles. The scattering matrix can be written in the following form
\bea
S_{\alpha ',\beta '} ^{\alpha \beta}(\lambda_1 - \lambda_2) &=&
\left[1+(1+\varepsilon_1)\cosh\gamma\Phi(\lambda_1 - \lambda_2)\right]
 \delta_{\alpha,\beta '} \delta_{\beta,\alpha '} \nonumber\\
& - & {\varepsilon_1} u_{\alpha}u_{\beta'}
\Phi (\lambda_1 - \lambda_2) \delta_{\alpha,-\beta}\delta_{\beta',-\alpha'},
 \eea
where
\bea
 \Phi(\lambda) = -\frac{\sin\lambda}{\sin(\lambda - i\gamma)}
\eea
and  $\lambda_j$ ($j=1,2,...,n$)  are suitable particle rapidities 
related to the momenta $\{k_j\}$ of the electrons by
\bea
k_j=\cases{
\pi-\Theta(\lambda_j;\frac{1}{2}\gamma),& ${\varepsilon\varepsilon_1}=-1$,\cr
-\Theta(\lambda_j;\frac{1}{2}\gamma),& ${\varepsilon\varepsilon_1}=+1$,\cr}
\eea
with the function $\Theta$  defined by
\bea
\Theta(\lambda;\gamma)=2\arctan\left(\cot\gamma\cdot\tan 
\lambda \right) ;\hspace{1cm}
-\pi<\Theta(\lambda,\gamma)\leq\pi.
\eea
A necessary and sufficient condition for the 
applicability of the Bethe-ansatz method is the Yang-Baxter 
equation [18,21]. In our case the S-matrix satisfies these 
equations in the non-deformed and q-deformed cases  [17]. The 
isotropic case corresponds for $S >\frac{1}{2}$ to the $q$-deformed case 
where $u_s = 1 (s = -S, \ldots, S)$ and $q + 1/q = 2S + 1$. The underlying 
Hecke algebra of the model implies that differently from the supersymmetric 
t-J model we should have gapped spin excitations for $S \geq 1$.  
  Up to our 
knowledge this model is the first example of integrable 
model with the S-matrix of the form (4) which is connected with the Hecke 
algebra.
 The Hamiltonian (3) is diagonalized by standard procedure by 
imposing periodic boundary conditions on the Bethe function.  
 These boundary 
conditions can be expressed in terms of the transfer matrix of the 
non-uniform model which can be constructed on the basis of the S-matrix (4) by 
using the quantum method of the inverse problem [22, 23]. The rapidities 
$\{\lambda_j\}$ that define a n-particle wave function are obtained by solving 
the equations
\bea
\left[\frac{\sinh(\lambda_j-i\gamma/2)}{\sinh(\lambda_j+
i\gamma/2)}\right]^L=(-1)^{n-1}\Lambda(\lambda_j),
\eea
where $\Lambda(\lambda)$ is the eigenvalue of the transfer matrix
\bea
T_{\{\alpha_l'\}}^{\{\alpha_l\}}(\lambda)=\sum_{\{\beta_l\}}\prod_{l=1}^n
S_{\alpha_l'\beta_l}^{\alpha_l\beta_{l+1}}(
\lambda_l-\lambda), \;\;\; (\beta_{n+1} = \beta_{1}).
\eea
It is simple to verify that besides the number of particles $n$, the 
magnetization $\sum_j S_j^z$ and the number of paired electrons $m$ 
 are conserved 
quantities in the Hamiltonian (3). Two electrons are paired if they 
are consecutive electrons with opposite spins 
and have no unpaired electron between them.
The complete diagonalization of the transfer matrix (9) is not a simple 
problem even in the simplest case $S=1, n = L$ (see, for example, [24]). 
It is not difficult to convince ourselves that in the interesting physical 
situation where we have low density of holes the ground-state will belong 
to the sector where we have zero magnetization and only pairs of electrons. In 
this sector $m =n/2$ and the diagonalization of the transfer matrix of the 
inhomogeneous model (9) gives for $\varepsilon_1 = -1$, the following equations 
\bea
\left[\frac{\sin(\lambda_j+i\gamma/2)}{\sin(\lambda_j-i\gamma/2)}\right]^L=(-1)^
{m-1}\prod_{\alpha=1}^m\frac{\sin(\lambda_j-\Lambda_{\alpha}+i\gamma/2)}
{\sin(\lambda_j-\Lambda_{\alpha}-i\gamma/2)} , \nonumber \\ 
\prod_{j=1}^n\frac{\sin(\lambda_j-\Lambda_{\alpha}-i\gamma/2)}
{\sin(\lambda_j-\Lambda_{\alpha}+i\gamma/2)} =-\prod_{\beta=1}^m\frac
{\sin(\Lambda_{\alpha}-\Lambda_{\beta}+i\gamma)}
{\sin(\Lambda_{\alpha}-\Lambda_{\beta}-i\gamma)} .
\eea
In the case $\varepsilon_1=+1$ the first set of equations in (10) should be 
replaced by 
\bea
\left[\frac{\sin(\lambda_j+i\gamma/2)}{\sin(\lambda_j-i\gamma/2)}\right]^L=(-1)^
{m-1}\prod_{l=1}^n\frac{\sin(\lambda_j-\lambda_l+i\gamma)}
{\sin(\lambda_j-\lambda_l-i\gamma)}
\prod_{\alpha=1}^m\frac{\sin(\lambda_j-\Lambda_{\alpha}-i\gamma/2)}
{\sin(\lambda_j-\Lambda_{\alpha}+i\gamma/2)}  \nonumber
\eea
The total energy and momentum of the model are given in 
terms of the particle rapidities $\lambda_j$ in the following form
\bea
E&=&-2\sum_{j=1}^n\cos k_j =
2{\varepsilon\varepsilon_1}\sum_{j=1}^n\left(\cosh\gamma-\frac{\sinh^2\gamma}
{\cosh\gamma-\cos2\lambda_j}\right), \nonumber\\
P&=&\sum_{j=1}^nk(\lambda_j) .
\eea
The equations (10) and (11) have the same structure as 
 those appearing in the anisotropic t-J model [15,16] provided a 
suitable definition of the parameter $\gamma$ is given. It means that in 
spite of the physical processes in the models with $S = \frac{1}{2}$ and 
$S > \frac{1}{2}$ being  
quite different there is a  "weak equivalence" in Baxter's sense
[25]  between models with different values of spin $S$ in the sector 
where $m=n/2$.
Of course in the general case this equivalence does not exist.

Although the models are exactly integrable for both signs of 
$\varepsilon$ and $\varepsilon_1$ in (3) let us now restrict to the 
more physically interesting 
case $\varepsilon = 1$ and $\varepsilon_1 = -1$, where we have attraction 
among pairs. In this case 
the ground state contains 
 $m=n/2$ bound pairs characterized 
 by a pair of complex electron rapidities

\bea
\lambda_{\alpha}^{\pm}=\frac{1}{2}(v_{\alpha}\pm i\gamma), 
 \;\; v_{\alpha}=2\Lambda_{\alpha}.
\eea
The second set of equation in (10) is fullfilled within 
exponential accuracy whereas the first set  can be 
treated in the similar way as in [15,16]. Inserting (12) in the first set 
of equations in (10) and introducing the density function $\rho(v)$ for the 
distribution of $v_{\alpha}$ in the thermodynamic limit, we obtain the 
linear integral equation
\bea
2\pi\rho(v)=\Phi(v;\gamma)-\int_I\Phi(v-v';\gamma)\rho(v')dv'
\eea
where
\bea
\Phi(v;\gamma)=\frac{\sinh 2\gamma}{\cosh 2\gamma-\cosh v}.
\eea
In order to minimize the ground-state energy
\bea
\frac{E_0}{L}= - 2\varepsilon\int_I\left[2\cosh\gamma - 
   \sinh\gamma\Phi(v;\gamma)\right]\rho(v)dv
\eea
the integration interval $I$ in (13) and  (15) 
has to be chosen symmetrically around
$\pi(I = [v_0, 2\pi - v_0]$. The parameter $v_0$ is 
determined by the subsidiary condition for the total 
density $\rho = 2m/L$ of  electrons
\bea
\int_I\rho(v)dv = \frac {1}{2}\rho .
\eea

To study the superconducting properties of the model under 
consideration we  calculate the long-distance behaviour of the 
correlation functions by  finite size studies and 
application of conformal field theory (see [26-28] and
references therein). The results of this calculation are the following. The
long-distance behavior of the density-density and the superconducting
correlation functions are given by

\bea
\left<\rho(r)\rho(0)\right>\simeq\rho^2+A_1r^2+A_2r^{-\alpha}
\cos(2k_Fr); \hspace{1cm}     2k_F=\pi\rho;
\eea
\bea
\rho(r)&=&\sum_{\gamma}c_{r\gamma}^+c_{r\gamma}, \nonumber \\
G_{\rho}(r)&=&\left<c_{r\gamma}^+c_{r+1,-\gamma}^+c_{0,\delta}c_{1,-\delta}
\right>\simeq Br^{-\beta} .
\eea
The exponents $\alpha$ and $\beta$ describing the algebraic 
decay are calculated from the dressed charge function $\xi(v)$ which is 
given by the solution of the integral equation
\bea
\xi(v)=1-\frac{1}{2\pi}\int_I\Phi(v-v';\gamma)\xi(v')dv',
\eea
and is given by 
\bea
\alpha=\beta^{-1}=2[\xi(v_0)]^2 .
\eea

In our one-dimensional system we have no superconductivity in the 
literal sense, since the model does not have finite off-diagonal 
long-range
 order. But we may say that in our model there is tendency to the 
superconductivity since  the superconducting correlations have a longer 
range than the density-density correlations. This happens when $\beta < 
\alpha$. Analytically we find $\alpha=2$ for ($\rho = 0$) and 
$\alpha =\frac{1}{2}$ for ($\rho = \rho_{max} = 1$). 
 This implies that for all 
nonzero values of the parameters $\gamma$ there is a density 
regime $[0,\rho_c]$
where the system has dominating superconducting 
correlations.
An  analogous behaviour of correlation functions can also be 
observed in the $SU(N)$
generalization of the anisotropic t-J model where 
superconducting properties are caused by the introduction of  
anisotropy in the  interactions. However unlike these models the 
superconducting properties in the Hamiltonians (3)  are 
caused by both effects, the anisotropy and the value of the spin S 
(see definition of 
the parameter $\gamma$ (3)). Moreover  in the present model for any 
value of $N$ ($N = 2S + 1$) we have bound pairs but not complexes of $N$ bound 
particles as in [16].

We conclude the letter with some remarks about the lattice vertex model 
counterpart of the quantum chain considered here.
The quantum $R$-matrix has $1+3N + 2N^2$ non-zero Boltzmann 
weights , which are given  by
\bea
R_{00}^{00} = 1, \; \; R_{0\alpha}^{0\alpha} = R_{\alpha 0}^{\alpha 0} = 
\varepsilon \sinh 
\lambda/\sinh(\gamma -\varepsilon_1 \lambda) \nonumber \\
R_{\alpha 0}^{0 \alpha} = R_{0\alpha}^{\alpha 0} = 
\sinh\gamma/\sinh(\gamma -\varepsilon_1 \lambda) \nonumber\\
R_{\gamma\delta}^{\alpha \beta} = 
\left[\delta_{\alpha,\delta}\delta_{\beta,\gamma} +
\Phi(i\lambda)u_{\alpha}
u_{\delta}
\delta_{\alpha,N-\beta +1}\delta_{\delta,N-\gamma +1}\right]
\sinh(\gamma -\lambda)/\sinh(\gamma -\varepsilon_1 \lambda) .
\eea
where $\alpha, \beta = 1,2,...,N $. The associated spin Hamiltonian can be 
found by taking the logarithmic derivative of the row-to-row 
transfer matrix at $\lambda = 0$. It
gives the Hamiltonian  (3) after a Jordan-Wigner 
transformation. Since we verified that (21) 
satisfy the Yang-Baxter equations 
the exact integrability of (3) is an immediate consequence.
 The above vertex  model  can be 
treated by the diagonal-to-diagonal Bethe ansatz 
method [29,30], but this is not the aim  of this letter.
\newpage
\begin{center}
{\bf Acknowledgments}
\end{center}
This work was supported in part by Conselho Nacional de Desenvolvimento 
Cient\'{\i}fico - CNPq - Brazil, by Funda\c c\~ao de Amparo \`a Pesquisa 
do Estado de S\~ao Paulo - FAPESP - Brazil, and by Russian Foundation of  Fundamental Investigations under Grant No.RFFI 97 - 02 - 16146. We would like
to thank Dr.H.Babujian for discussions.

\newpage
{\bf REFERENCES}
\begin{enumerate}
\item F. C. Zhang and T.M.Rice, Phys. Rev. B {\bf 37}, 3759 (1988).
\item P. W. Anderson, Phys. Rev. Lett. {\bf 65}, 2306 (1990).
\item B. Sutherland, Phys. Rev. B {\bf 12} 3795 (1975).
\item M. Jimbo, Lett. Math. Phys. {\bf 11} 247 (1986).
\item P. Schlottmann, Phys. Rev. B {\bf 36}, 5177 (1987).
\item P. A. Bares and G. Blatter, Phys. Rev. Lett. {\bf 64}, 2567 (1990).
\item N. Kawakami, S.-K.Yang, Phys. Rev. Lett. {\bf 65} 2309 (1990).
\item K. Lee, P. Schlottmann, J. Phys. Colloq. {\bf 49} C8 709 (1988).
\item P. Schlottmann, J. Phys. C {\bf 4}, 7565 (1992).
\item N. Kawakami, Phys.Rev. B {\bf 47}, 2928 (1993).
\item F. H. L. Essler, V. E. Korepin and K. Schoutens, Phys.Rev.Lett. 
 {\bf 68}, 2960 (1992); 70, 73 (1993).
\item F. H. L. Essler and  V. E .Korepin, Exactly Solvable Models of Strongly
 Correlated Electrons (World Scientific, Singapore, 1994).
\item R. Z. Bariev, J. Phys. A {\bf 27}, 3381 (1994).
\item A. Forster and  M. Karowski, Nucl. Phys. B {\bf 408 [FS]}, 512 (1993).
\item R. Z. Bariev, Phys. Rev. B {\bf 49}, 1447 (1994).
\item R. Z. Bariev, A. Kl\"umper, A. Schadschneider and J. Zittartz,
Z. Phys. B {\bf 96}, 395 (1995).
\item F. C. Alcaraz, R. K\"oberle and A. Lima-Santos, Int. J. Mod. Phys. 
A {\bf 7}, 7615         (1992).
\item C. N. Yang, Phys. Rev. Lett. {\bf 19}, 1312 (1967).
\item  E. H. Lieb and F. Y. Wu, Phys. Rev. Lett. {\bf 20}, 1445 (1968).
\item  B. Sutherland, Phys. Rev. Lett. {\bf 19}, 103 (1967).
\item  R. J. Baxter, Exactly solved models in statistical mechanics
    (Academic Press, New York, 1982).
\item L. Takhtadzhyan and  L. D. Faddeev, 
Russ. Math. Survey {\bf 34}, 11 (1979).
\item  V. E. Korepin, A. G .Izergin and N. M. Bogoliubov, Quantum Inverse 
Scattering Method, Correlation Functions and Algebraic Bethe Ansatz 
(Cambridge University Press, Cambridge, 1993)
\item A.Kl\"umper, Europhys.Lett.{\bf 9}, 815 (1989);\\
      R. K\"oberle and  A. Lima-Santos, J.Phys. A {\bf 27}, 5409 (1994).
\item J. H. H. Perk and F. Y. Wu, Physica A 138 (1986) 100;
J. Stat. Phys. {\bf 42},  727 (1986).
\item J. L. Cardy, Nucl. Phys. B {\bf 270 [FS16]}, 186 (1986).
\item R. Z. Bariev, A. Kl\"umper, A. Schadschneider and
J. Zittartz, J. Phys. A {\bf 26}, 1249; 4863 (1993); Physica B
{\bf 194-196}, 1417 (1994).
\item N. M. Bogoliubov and  V. E. Korepin, Int. J. Mod. Phys. B {\bf 3}, 
427 (1989).
\item R. Z. Bariev, Theor. Math. Phys. 49 (1982) 1021;
\item   T. T. Truong, K. D. Sch\"otte, Nucl. Phys. B {\bf 220[FS8]} 77 (1983).

\end{enumerate}
\end{document}